\documentclass[aps,a4paper,twocolumn]{revtex4}

\newcommand{\be}{\begin{equation}}
\newcommand{\ee}{\end{equation}}
\newcommand{\ba}{\begin{array}}
\newcommand{\ea}{\end{array}}
\newcommand{\bea}{\begin{eqnarray}}
\newcommand{\eea}{\end{eqnarray}}
\newcommand{\bdm}{\begin{displaymath}}
\newcommand{\edm}{\end{displaymath}}

\begin{document}

\title{Random Field Ising Model} 
\author{Prabodh Shukla}\email{shukla@nehu.ac.in }  
\affiliation{ Physics Department, North Eastern Hill University
\\Shillong-793022, 
India.}

\begin{abstract} 

Recent work on random field Ising model is described briefly emphasising
exact solutions of the model in simple cases and their relevance in
understanding equilibrium and non-equilibrium properties of systems with
quenched disorder. 

\end{abstract}

\maketitle

\section{Introduction:}

Extended systems with quenched random disorder often possess a large
number of nearly degenerate states separated by high energy barriers. This
kind of energy landscape gives rise to very complex relaxation phenomena
at low temperatures as well as hysteresis in the system. Even a weak
disorder may modify the system significantly and destroy the long range
order in the equilibrium state of the system. One needs a simple
theoretical model to make sense of the rich, complex, and vast amount of
experimental data in this field. The random field Ising model is perhaps
the simplest minimal model that fits the bill. Although very simple to
state, the model is not exactly solvable except in a few special cases.
Computer simulations of the model are helpful, but are often unable to
resolve the key questions because they suffer from metastability and slow
relaxation rates in the system just as much as the laboratory experiments.
For specificity, we focus on disordered magnets. Some of the important
questions one would like to be answered are the following: What is the
lower critical dimensionality below which any disorder, no matter how
small, would destroy the long range order in the system at zero
temperature? Could the disorder cause a first order jump discontinuity in
the response of the system to an applied field ? If so, in what
circumstances? In recent years, we have examined these questions in some
exactly solved cases of the model. Due to the limitation of space, we only
indicate the scope and status of our studies, and mention the results
obtained so far. Interested readers should consult the references for more
technical details. 

\section{The Model}

The model is defined on a lattice. Each site is labeled by an integer i,
and carries an Ising spin $S_{i}$ ( $S_{i}= \pm 1$ ), a quenched random
magnetic field $h_{i}$, and an externally applied uniform field $h$. The
quenched fields $\{ h_{i} \}$ are independent identically distributed
random variables with a continuous probability distribution $\phi(h_{i})$.
For convenience, we assume $\phi(h_{i})$ to be Gaussian with mean value
zero, and variance $\sigma^2$. The nearest neighbor interaction can be
ferromagnetic (J$>$0) or anti-ferromagnetic (J$<$0). The Hamiltonian of
the system may be written as,

\be H=-J \sum_{i,j} S_{i} S_{j} - \sum_{i} h_{i} S_{i} - h \sum_{i} S_{i}
\ee

The spin $S_{i}$ experiences a net field $f_{i}$ on it that is given by,

\be f_{i}=J \sum_{j} S_{j} + h_{i} + h \ee

The Glauber dynamics of the system at temperature T is specified by the
rate $R_{i}$ at which a spin $S_{i}$ flips to $- S_{i}$

\be R_{i}=\frac{1}{\tau} \left[ 1 - S_{i} \mbox{ tanh}\{ f_{i}/(k_{B}T) \}
\right], \ee

Here $\tau$ sets the basic time scale for the relaxation of individual
spins. The energy of the spin $S_{i}$ is equal to $ - f_{i} S_{i} $. If $
f_{i}$ and $S_{i}$ have the same sign, we say that the spin is aligned
along the net field at its site. The energy of a spin is the lowest if it
is aligned along the net field at its sight. We are interested in the
properties of the model at zero temperature ( T = 0 ), and on time scales
much larger than $\tau$. In this limit, the dynamics simplifies to the
following rule: choose a spin at random, and flip it only if it is not
aligned along the net field at its site.  Repeat the process till all
spins are aligned along the net fields at their respective sites. The
dynamics described above always brings the system to a fixed point state
that is stable against single spin flips. The fixed point corresponds to a
local minimum of the energy of the system. The system possesses a
thermodynamically large number of fixed point states each marking a domain
of attraction in the phase space of the system. The particular local
minima that the system reaches depends on the domain that contained the
initial state of the system. 

Although the fixed points are stable, they can be thought to represent the
metastable states of the system in the sense that they are local minima of
energy. In this context, it is useful to consider the time scales of
interest in physical systems, and the appropriateness of the
zero-temperature dynamics as a model. In a typical experiment on
non-equilibrium behavior of a physical system, there are at least four
time scales:  (i) time $\tau$ that an individual spin takes to relax, (ii)
time $\tau_{1}$ that the system takes to relax to a metastable state,
(iii) time $\tau_{2}$ over which the applied field changes, and (iv) life
time $\tau_{3}$ of the metastable state. In complex systems, $\tau_{1}$,
$\tau_{2}$, and $\tau_{3}$ may each contain an entire spectrum of time
scales. In physical systems relevant to our model, the shortest time is
$\tau$ which may be taken to be unity to set the scale. The next larger
time is $\tau_{1}=\nu \times \tau$, where $\nu$ is the number of
iterations of the dynamics to reach a fixed point. The applied field is
assumed to vary very slowly (driving frequency goes to zero!) so that
$\tau_{2} >> \tau_{1}$. Specifically, it means that the applied field is
held constant during the relaxation of the system. The time $\tau_{3}$ is
infinite. Thus, the present model is applicable if $ \tau_{3} >> \tau_{2}
>> \tau_{1} >> \tau$. These conditions fit a wide class of complex
magnetic materials that have a large number of metastable states separated
from each other by barriers much larger than the available thermal energy. 

\section{Relevance to Experiments}

At an intuitive level, quenched random fields may be thought to arise from
impurities and imperfections in disordered magnetic materials. A more
precise connection between random field Ising model and experiments was
made by a theoretical argument due to Fishman and Aharony ~\cite{fishman}
suggesting that the critical behavior of a weakly diluted anti-ferromagnet
in a uniform external field should be in the same universality class as
that of a ferromagnet in a random external field. The essential idea is
that an anti-ferromagnet without dilution has two sub-lattices on which
the spins are oriented opposite to each other. One of the sub-lattice is
aligned with the applied field. In the presence of dilution, locally the
sub-lattice with most spins tends to align with the applied field in
competition with the global anti-ferromagnetic order in the absence of
dilution. As a result the applied field acts as an effective random field
coupling to the anti-ferromagnetic order parameter i.e. the staggered
magnetization. The effective random field produced by the applied field is
proportional to the applied field. The strength of the random field is
therefore easily controlled. One can do scaling studies with varying
strengths of disorder in a system by simply tuning the applied field
rather than making fresh samples with different degrees of dilution. This
feature has a far reaching significance in investigating the theoretical
model experimentally, and the interaction between theory and experiment
has helped the field develop considerably, and clarified many fine points
of the model ~\cite{belanger,nattermann}.  $Rb_{2}CoF_{4}$ is a good two
dimensional Ising anti-ferromagnet. It consists of layers of magnetic ions
with a single dominant intralayer exchange interaction, and an interlayer
interaction which is smaller by several orders of magnitude. It is very
anisotropic so that the spins can be well represented as Ising spins. The
material can be magnetically diluted by introducing a small fraction of
manganese ions in place of cobalt ions.  Thus crystals of
$Rb_{2}Co_{x}Mn_{1-x}F_{4}$ are good examples of a two dimensional diluted
anti-ferromagnet suitable for an experimental realization of the two
dimensional random field Ising model. In three dimensions, the most
studied dilute anti-ferromagnet is $Fe_{x}Zn_{1-x}F_{2}$ crystal. In the
pure ferrous fluoride $FeF_{2}$ crystal, the ferrous ions ( $Fe^{++}$ )
are situated approximately on a body centered tetragonal lattice. Each
ferrous ion is surrounded by a distorted octahedron of flurine ( $F^{-}$ )
ions. The predominant interactions are a large single-ion anisotropy, and
an anti-ferromagnetic exchange between nearest neighbor ferrous ions. The
magnetic moments of ferrous ions on the corners of the tetragonal cell are
anti-parallel to the magnetic moments of $Fe^{++}$ ions on the body
centers. The large crystal field anisotropy persists as the magnetic spins
are diluted with $Zn$. The diluted crystal remains an excellent Ising
anti-ferromagnet for all ranges of magnetic concentration $x$. Furthermore
crystals with excellent structural quality can be grown for all
concentrations $x$ with extremely small concentration variation $ \delta x
< 10^{-3}$. These attributes combine to make $Fe_{x}Zn_{1-x}F_{2}$ the
popular choice for experiments on diluted anti-ferromagnets, although
experiments have been done on several other materials as well
~\cite{belanger}. 

\section{Domains in Random Fields }

In one of the earliest studies of the random field Ising model, Imry and
Ma~\cite{imma} argued that quenched random fields in a system may cause a
uniform ferromagnetic state to break into domains. The argument is
essentially as follows. Consider a uniform ferromagnetic state at zero
temperature. In the presence of random fields, a strategically placed
domain of linear size L may turn over and gain an energy of the order of
$\sigma L^{d/2}$.. However, this would create a domain wall that would
cost an energy of the order of J$L^{d-1}$. If $ d/2 > (d-1) $, i.e. if $ d
< 2 $, then for any $\sigma$, there will be a characteristic length over
which the bulk energy gain will overcome the cost of the surface energy.
In other words, domains will occur spontaneously if $ d < 2 $. The above
argument is intuitively appealing, but nonrigorous. It is also
inconclusive at the lower critical dimensionality because the gain and the
cost of energy both scale linearly with $L$ at $d=2$ . The situation in
two dimensions can be clarified by taking into account the roughness of
the domain wall. The work of Binder ~\cite{binder} supplemented with
numerical simulation of a toy model predicts that the gain in energy
scales as $\sigma L \log L$ in $d=2$. Thus the domain argument predicts
the absence of long range order in two dimensions. 

It remained unclear for several years if the results obtained from the
domain argument were correct. The controversy was generated by the
existence of another argument based on a field theoretic method
(dimensional reduction) that predicted the lower critical dimensionality
of random field Ising model to be three. Initially, computer simulations
of the model as well as experimental observations did not help to resolve
the controversy. It took several years to realize the error in the
dimensional reduction argument. The difficulty with both theory and
experiment was in the interpretation of results. Concentration gradients
in the diluted anti-ferromagnetic samples tend to round off a transition
and affect the measurements of critical behavior drastically. Further,
the majority of experiments were performed on samples prepared in two
separate ways; (i) cooling the sample in zero magnetic field, and (ii)
cooling it in a magnetic field, and turning the field off at the end.
Experiments on samples prepared in the two ways yield different results.
Three dimensional field cooled samples show no long range order and were
first thought to show that the lower critical dimensionality of the model
is three. But three dimensional samples cooled in zero field showed long
range order. It was only after several years of controversy that the
experimental situation resolved itself in favor of the domain argument,
i.e. no long range order in two dimensions, but long range order in three
dimensions. The main point that was clarified by theoretical work is that
the field cooled state is not an equilibrium state. It relaxes
logarithmically slowly, and one should not expect to see an equilibrium
ordered state in a field cooled sample over any reasonable experimental
time scale. 

\section{Hysteresis}

Permanent magnets are typically a two phase solid material with fine
magnetic particles of one phase embedded in the other phase. The
precipitation is carried out in a magnetic field, and needle like magnetic
particles are oriented with their long axis parallel to the field
direction. If the field is reversed after the material is set in a solid
matrix, the needles can not physically turn over as in a liquid matrix.
Instead, the magnetic domains inside the particles have to reverse
themselves. This requires a threshold field that varies from particle to
particle. Experimentally observed magnetization of the material in a
smoothly increasing applied field may look smooth on a macroscopic scale
but on a microscopic scale, it is made of steps of irregular widths and
heights. This is known as Barkhausen noise.  Hysteresis in such materials
is somewhat different in character from the usual hysteresis that arises
from a delayed response of the system to the driving field. In the present
case the hysteresis is dominated by the energy landscape of the system.
There are a very large number of local minima with barriers of varying
heights between them. Suppose the applied field is increased sufficiently
so that the system jmps over a barrier from one minimum A to another
minimum B. If the field is then reversed, it may be more favorable for the
system to jump over a smaller barrier to another minimum C rather than
return to A. This is the kind of hysteresis that is modeled very well by
the zero temperature dynamics of the random field Ising model
~\cite{sethna}. We have solved the ferromagnetic model exactly on a Bethe
lattice of an arbitrary coordination number $z$, and obtained major as
well as minor hysteresis loops, and power laws for the Barkhausen noise
analytically \cite{shukla1,dhar,sabha1,shukla2,shukla3,sabha2}. In the
following, we describe our main results without going into the details of
calculation. 

The main limitation of the method used to obtain theoretical hysteresis
loops has been its restriction to Bethe lattices, and to initial states
having saturated magnetization i.e. all spins pointing in the same
direction. More recently the method has been extented to the case of a
random initial state ~\cite{shukla5} but we shall not go into it here.
Suppose all the spins are pointing down.  This is a stable state in an
applied field $h=-\infty$. We need to calculate the magnetization as the
applied field in increased from $h=-\infty$ to $h=\infty$. The other half
of the hysteresis loop can be obtained by a symmetry of the Ising model,
and that of the Gaussian distribution centered at the origin. In the first
instance, we calculated the magnetization on a one dimensional lattice
~\cite{shukla1}. The calculation utilized the fact that the evolution on a
linear segment of the lattice contained between two sites that flip up
before their nearest neighbors is independent of the rest of the lattice.
It was not immediately obvious how the calculation could be generalized to
a Bethe lattice. What makes the generalization possible is the following:
Suppose the spin $S_{i}$ in the deep interior of the Bethe lattice does
not flip up as the applied field increases from $h-\infty$ to $h$, then
the evolution on each of the $z$ sublattices that meet at site {i} is
independent of each other upto the applied field $h$.  The second point is
that the stable state at applied field $h$ does not depend on the order in
which the spins are relaxed if the interactions are ferromagnetic. This
property of the ferromagnetic model is called the Abelian property. It
allows us to relax the nearest neighbors of site $i$ independently of each
other and before site $i$ is relaxed. The method involves calculating the
conditional probability that a nearest neighbor of site $S_i$ is up before
$S_i$ flips up. This conditional probability is obtained as a fixed point
of a recursion relation describing the relaxation of sites starting from
the boundary of the Bethe lattice and working in. 

Hysteresis in the ferromagnetic model on a Bethe lattice has symmetrically
placed jump discontinuities on the two halves of the hysteresis loop if
$\sigma$ is smaller than a critical value $\sigma_c$, and the coordination
number $z$ of the lattice is greater than or equal to four. In other
words, hysteresis loops on a Bethe lattice of coordination z $>$ 3 are
qualitatively different from those on lattices with z=2, and z=3. For $ z
\le 3$ there is no jump discontinuity in the hysteresis loops for any
amount of disorder. For $ z \ge 4$, there is a critical value of $\sigma$
that characterizes the Gaussian random field distribution. If $\sigma$ is
less than the critical value $\sigma_{c}$, the magnetization in increasing
field has a macroscopic first-order jump at an applied field $h_{c}$ $>$
J. As $\sigma$ increases to $\sigma_{c}$, $h_{c}$ decreases to J, and the
first-order jump in magnetization reduces to zero. The system shows
non-equilibrium critical behavior at $h=h_{c}$, and $\sigma=\sigma_{c}$.
For z=4, $\sigma_{c}=1.78$ approximately. The value of critical disorder
$\sigma_{c}$ increases with the coordination number $z$ of the lattice. 
At the critical point the Barkhausen noise shows a power law distribution. 
The probability of avalanches of size $s$ scales as $s^{-3/2}$.
Nonequilibrium critical point phenomena and its relationship with the
coordination number of the lattice appears to hold on other lattices as
well. Numerical simulations and theoretical arguments on several periodic
lattices embedded in two and three-dimensional space show that hysteresis
on periodic lattices with $z \ge 4$ is qualitatively different from that
on lattices with $z < 4$. Although there are some similarities between
Bethe lattices and periodic lattices of the same coordination number,
there are differences as well. The differences are related to the
bootstrap percolation instability on some periodic lattices
~\cite{sabha3}. 

Minor hysteresis loops have also been obtained, and their calculation
reveals an interesting feature of the model that has an experimental
significance. Consider two halves of the major hysteresis loop connecting
states of saturated magnetization in opposite directions. If the applied
field is reversed while the system is on one of these halves, the
magnetization trajectory branches off and heads towards the other half of
the major loop. We find that the trajectory in reversed field meets the
other half of the major loop exactly when the field has been reversed by
an amount 2J irrespective of the point of reversal. This result provides
an interesting possibility for measuring the exchange interaction J in a
hysteresis experiment.

We have also studied hysteresis in the anti-ferromagnetic random field
Ising model at zero temperature ~\cite{shukla4}.  Hysteresis in the
anti-ferromagnetic model is qualitatively different from that in the
ferromagnetic model because it does not show Barkhausen noise.  On account
of the anti-ferromagnetic interactions, a spin turning up in an increasing
field blocks its nearest neighbors from turning up.  Thus there is no
microscopic avalanche of up-turned spins as in the case of ferromagnetic
interactions. A spin that turns up in increasing field in the
anti-ferromagnetic model, occasionally causes its nearest neighbor (that
had turned up earlier) to turn down. As the applied field increases from
$-\infty$ to $+\infty$, a small fractions of sites flip three times, first
up, then down, and finally up again. Also, the dynamics of the
anti-ferromagnetic model is non-Abelian. These features make the
anti-ferromagnetic dynamics rather complex, and an exact solution of the
model becomes difficult. So far the anti-ferromagnetic model had been
solved exactly in one dimension only, and that too for a rectangular
distribution of the random field of width $2\Delta$, where $\Delta \le
|J|$. Recently, we have been able to extend the calculation of hysteresis
in a one dimensional random field Ising model to an arbitrary continuous
distribution of the random field ~\cite{shukla5}.

\end{document}